\documentclass[12pt, preprint]{aastex}
\usepackage{natbib}
\newcommand{\farc}{\hbox{$.\!\!^{\prime\prime}$}} 
\newcommand{\griz}{$g\arcmin r\arcmin i\arcmin z\arcmin$~}
\newcommand{\gri}{$g\arcmin r\arcmin i\arcmin$~}
\newcommand{\JHK}{$JHK_S$~}
\begin{document}


\title{Correlated optical and X-ray flares in the afterglow of XRF 071031}

\author{T. Kr\"{u}hler\altaffilmark{1,2}, J. Greiner\altaffilmark{1}, S. McBreen\altaffilmark{1,3}, S.Klose\altaffilmark{4}, A.Rossi\altaffilmark{4}, P. Afonso\altaffilmark{1}, C. Clemens\altaffilmark{1}, R. Filgas\altaffilmark{1}, A. K\"{u}pc\"{u} Yolda\c{s}\altaffilmark{5}, G. P. Szokoly\altaffilmark{6}, A. Yolda\c{s}\altaffilmark{1} } 

\altaffiltext{1}{Max-Planck-Institut f\"{u}r extraterrestrische Physik, Giessenbachstra\ss e, 85748 Garching, Germany; kruehler@mpe.mpg.de}
\altaffiltext{2}{Universe Cluster, Technische Universit\"{a}t M\"{u}nchen, Boltzmannstra\ss e 2, 85748, Garching, Germany}
\altaffiltext{3}{School of Physics, University College Dublin, Dublin 4, Ireland}
\altaffiltext{4}{Th\"{u}ringer Landessternwarte Tautenburg, Sternwarte 5, 07778 Tautenburg, Germany}
\altaffiltext{5}{European Southern Observatory, Karl-Schwarzschild-Stra\ss e 2, 85748 Garching, Germany}
\altaffiltext{6}{Institute of Physics, E\"{o}tv\"{o}s University, P\'{a}zm\'{a}ny P. s. 1/A, 1117 Budapest, Hungary}


\begin{abstract}

We present a densely sampled early light curve of the optical/near-infrared (NIR) afterglow of the X-Ray Flash (XRF) 071031 at $z$=2.692. Simultaneous and continuous observations in seven photometric bands from $g\arcmin$ to $K_S$ with GROND at the 2.2~m MPI/ESO telescope on LaSilla were performed between 4 minutes and 7 hours after the burst. The light curve consists of 547 individual points which allows us to study the early evolution of the optical transient associated with XRF 071031 in great detail. The optical/NIR light curve is dominated by an early increase in brightness which can be attributed to the apparent onset of the forward shock emission. There are several bumps which are superimposed onto the overall rise and decay. Significant flaring is also visible in the \textit{Swift} X-Ray Telescope (XRT) light curve from early to late times. The availability of high quality, broadband data enables detailed studies of the connection between the X-ray and optical/NIR afterglow and its colour evolution during the first night post burst. We find evidence of spectral hardening in the optical bands contemporaneous with the emergence of the bumps from an underlying afterglow component. The bumps in the optical/NIR light curve can be associated with flares in the X-ray regime suggesting late central engine activity as the common origin. 

\end{abstract}


\section{Introduction}

Major progress in the understanding of the X-ray and optical afterglow light curves of Gamma-Ray Bursts (GRBs) and the softer X-Ray Flashes (XRFs) has been made since the launch of the \textit{Swift} satellite \citep{geh04} and the rapid follow-up data provided by the X-Ray (XRT; \citealp{bur05}) and Ultra-Violet Optical Telescope (UVOT; \citealp{rom05}). However, access to the longer wavelength afterglow is still somewhat limited to the brighter half of all detected bursts. In contrast to the evidence of a generic X-ray afterglow light curve \citep{nou06}, the few bursts with very early detected optical counterparts show considerable variety. For instance GRBs 990123 \citep{ake99} and 041219A \citep{bla05, ves05, smb06}, have shown optical emission contemporaneous with the prompt phase of the burst. A significant delay of the apparent onset of the afterglow forward shock (FS) allowed ground based optical/NIR telescopes to detect a rising component of the afterglows for e.g. GRBs 030418 \citep{ryk04}, 060418, 060607A \citep{mol07}, 070802 \citep{kru08} and 071010A \citep{cov08}. A number of optical afterglows showed bumps superimposed onto the overall power law decay in late epochs, which are generally interpreted as the signature of either inhomogeneities in the circumburst medium (e.g. GRB 050502A, \citealp{gui05}) or late energy injections (e.g. GRB 021004, \citealp{uga05} or GRB 070311, \citealp{gui07}). 

The very early optical afterglow is of significant interest from a theoretical point of view and in particular how it relates to the flares and plateaus seen in many X-ray afterglow light curves \citep[e.g.][]{bri06}. In the early phase the colour evolution is crucial to differentiate between different emission components. However, most of the rapid ground based follow up is obtained with robotic telescopes of small aperture size in white light or filter cycles. In both cases information about the spectral properties is absent or can only be obtained at relatively long times with respect to the dynamical time scale in the early evolution of GRB emission. The ambiguity between effects of a changing spectrum or a highly variable early light curve can only be addressed by systematic observations in different broad-band filters as synchronous and rapid as possible. Comprehensive data sets of early optical afterglows were published e.g. for GRB~021004 \citep{laz02}, GRB~030329 \citep{lip04}, GRB~061126 \citep{per07} and the very bright GRB~080319B \citep[e.g.][]{rac08} where the light curve is well sampled in time and frequency domains, suggesting that a standard jet break model alone can not account for the increasing variety of features in a GRB or XRF afterglow.
 
Here we report on the optical follow up of GRB~071031 at redshift 2.692 \citep{led07} using data obtained in seven broad-band filters from $g\arcmin$ to $K_S$ with the multi-channel imager GROND \citep{gre07, gre08}. Ground-based optical/NIR observations started at $\sim$4~minutes after trigger, yielding one of the best sampled early optical light curves. In combination with the detailed X-ray observations provided by the XRT, this constitutes a multi color light curve with spectral coverage from the near-infrared (NIR) to the 10 keV XRT band.

\section{Observations}
\label{obs}
\subsection{\textit{Swift}}

The Burst Alert Telescope (BAT, \citealp{bar05}) on-board the \textit{Swift} satellite triggered on the long-soft GRB~071031 at T$_{\rm 0}$=01:06:36~UTC and immediately slewed to the burst \citep{str07}. 
The BAT light curve shows a two-peaked structure starting at T$_{\rm 0}$-10 s and ending at T$_{\rm 0}$+180s with a T$_{\rm 90}$ of 180$\pm$10~s. The fluence in the 15~keV to 150~keV band is 9.0$\pm$1.3 $\times$ 10$^{-7}$ erg cm$^{-2}$, with a fluence ratio of 1.34 between the BAT 25-50~keV and 50-100~keV bands \citep{sta07}. This is remarkably soft compared to conventional GRBs and qualifies GRB~071031 as a XRF according to the working definition of \citet{sak08}. The BAT spectrum of the first peak is well described with a single power law with a photon index of 2.26$\pm$0.30 ($\chi^2$=28.83 for 36 degrees of freedom). This is well outside the normal range for the low energy index $\alpha$ but similar to the spectral index $\beta$ above the break energy for a Band function \citep{ban93, pre00}. The peak energy of the prompt emission spectrum must then be close to or below the BAT lower energy range of around 30~keV (see also \citealp{mcg05, sta07}). Therefore GRB~071031 is designated as XRF~071031 hereafter.

\begin{figure}
\includegraphics[angle=270, width=0.98\columnwidth]{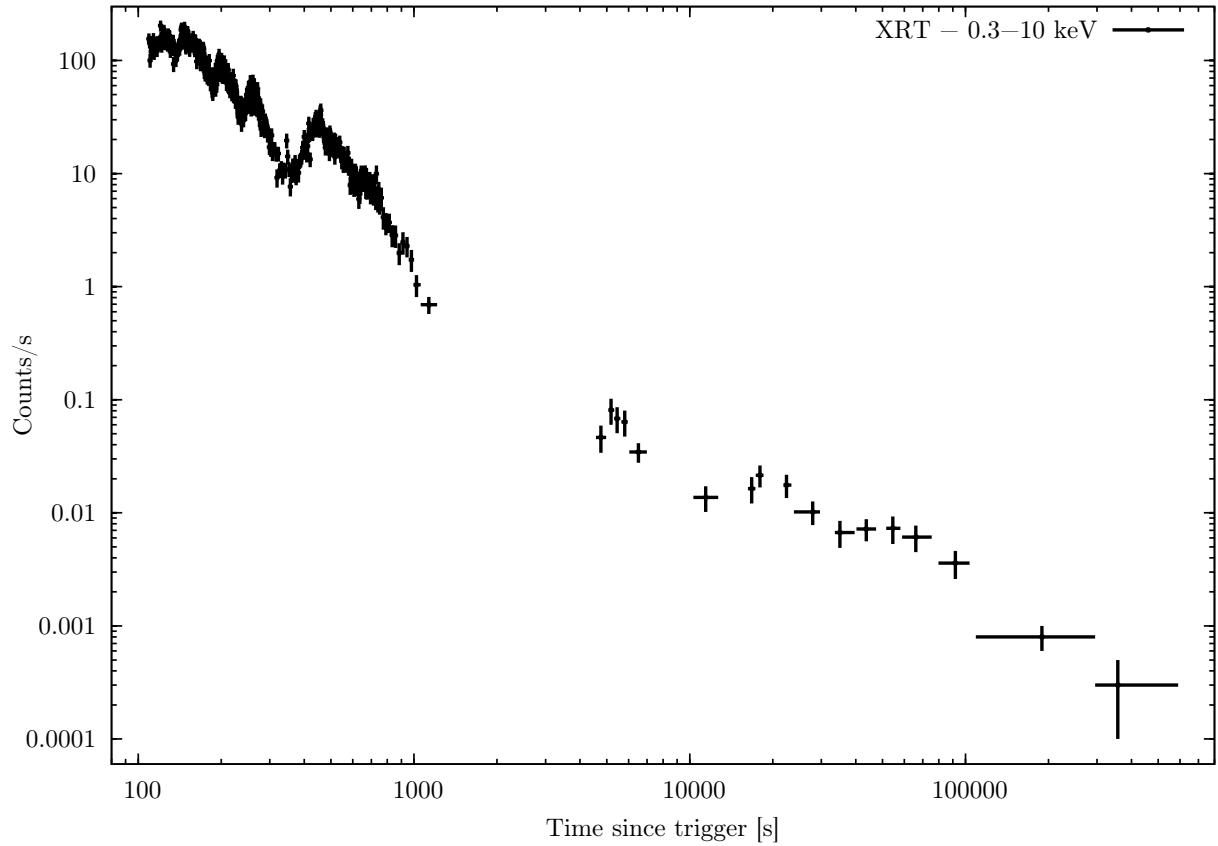}
\caption{\textit{Swift} XRT light curve of the X-ray afterglow of XRF~071031 obtained from the XRT light curve repository \citep{eva07}}
\label{picXRT}
\end{figure}

The XRT began follow-up observations of the burst field 103~s after the trigger and detected an uncatalogued fading X-ray source at a position of RA(J2000)=00$^h$~25$^{m}$~37$^s$.4, Decl(J2000)=-58$\arcdeg$~03$\arcmin$~33$\arcsec$ with a refined 90\% confidence error circle of 2\farc0 radius \citep{str07a}. The early XRT light curve is dominated by significant amount of flaring with bright flares at around 120~s, 150~s, 200~s, 250~s and 450~s. Also the late X-ray data exhibit rebrightenings at 5.5~ks, 20~ks and 55 ks superimposed onto the overall power law decay. The complete XRT light curve is shown in Fig.~\ref{picXRT}.

The X-ray spectra were obtained with the \texttt{xrtpipeline} tool using the latest calibration frames from the \textit{Swift} CALDB and standard parameters. The spectra were fitted with the XSPEC package \citep{arn96} and a foreground hydrogen column density at the Galactic value of $N_{\rm H}$=1.2$\times 10^{20}$~cm$^{-2}$ \citep{kal05}.

The third instrument on-board \textit{Swift}, UVOT started observations at T$_{\rm 0}$+114~s and found a transient source inside the XRT error circle in the white, $v$- and $b$-band filters. The UVOT data show an increase in the brightness of the afterglow of around 0.5~mag in the first few hundred seconds \citep{bre07}.

\subsection{GROND}

GROND responded to the \textit{Swift} GRB alert and initiated automated observations that started at 01:10:21 UTC, 3~minutes 45~s after the burst and continued until local Sunrise at 08:55:51 UTC. A predefined sequence of observations with successively increasing exposure times was executed and images were acquired in all seven photometric bands simultaneously. In total 84 individual frames in each \griz and 1510 images of 10~s exposures in \JHK  were obtained during the first night at airmasses between 1.1 and 2.4. The integration time of the CCD optical images scaled from 45~s to 360~s according to the brightness of the optical afterglow. A variable point source was detected in all bands \citep{kru07} by the automated GROND pipeline \citep{aky07} and its absolute position is measured to RA(J2000)=00$^h$~25$^{m}$~37$^s$.24, Dec(J2000)=--58$\arcdeg$~03$\arcmin$~33\farc6 compared to USNO-B reference field stars \citep{mon03} with an astrometric uncertainty of 0\farc3. Photometry and spectroscopy of the afterglow was also obtained by telescopes at CTIO \citep{hai07, cob07} and the VLT, the latter yielding an UVES and FORS spectroscopic redshift of 2.692 \citep{led07, fox08}.


\begin{figure}
\epsscale{0.95}
\plotone{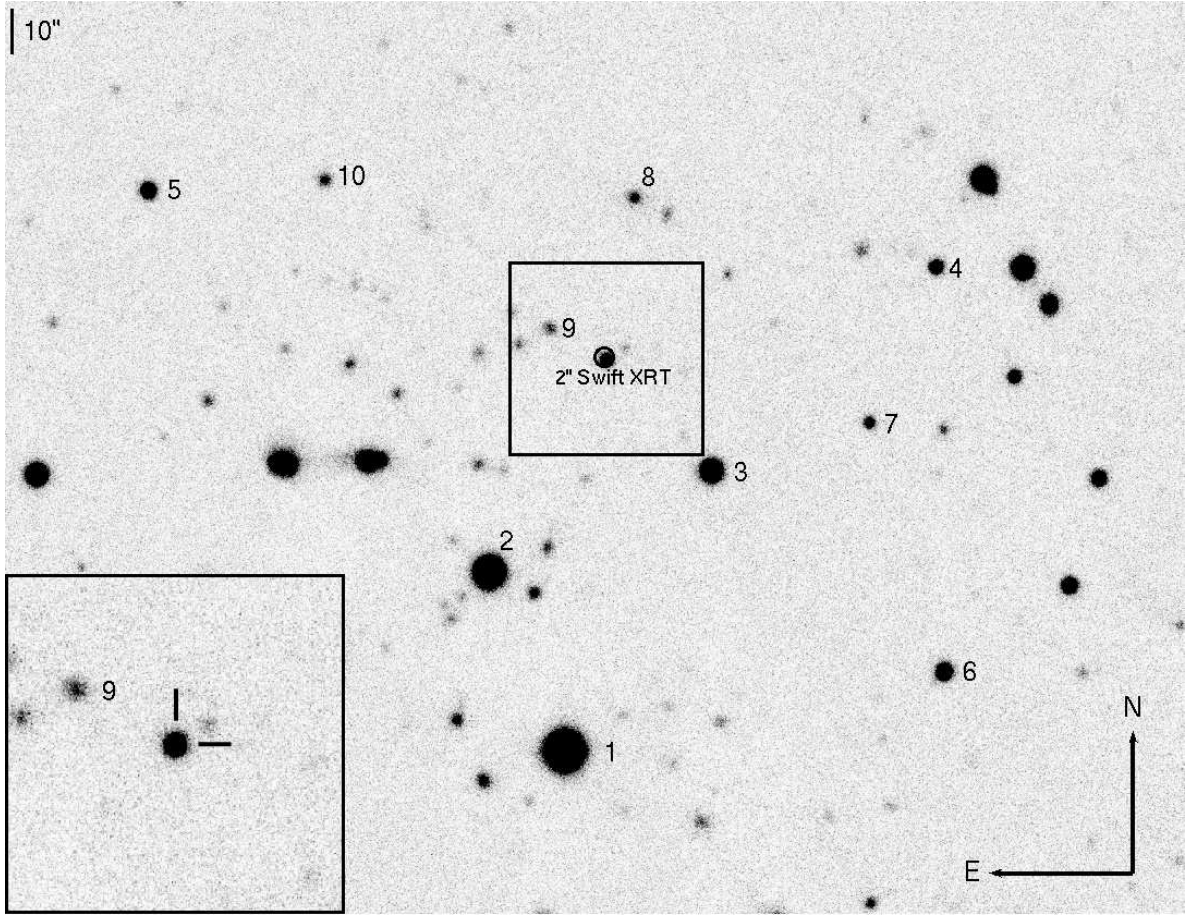}
\caption{GROND r$\arcmin$ band image showing the afterglow of XRF~071031 and the XRT error circle. The secondary standards used for calibration are labeled 1 to 10 and listed in Tab~.\ref{tabSecStan}. The lower left corner shows a zoom in to the afterglow position.}
\label{picStar}
\end{figure}

Photometric calibration was performed relative to the magnitudes of ten secondary standards in the field of XRF~071031, shown in Fig.~\ref{picStar} and Tab.~\ref{tabSecStan}. During photometric conditions, three spectrophotometric standard stars, SA114-750, SA114-656 and SA95-42, all primary Sloan standards \citep{smi02}, were observed with GROND. Observations of the GRB field followed within few minutes. The magnitudes of the Sloan standards were transformed to the GROND filter system using their spectra and the GROND filter curves \citep{gre08}. The obtained zeropoints were corrected for atmospheric extinction differences and used to calibrate the stars in the GRB field. An independent absolute calibration was obtained with respect to magnitudes of the SDSS and 2MASS stars within the standard fields obtained from the SDSS data release 6 \citep{ade08} and the 2MASS catalogue \citep{skr06} with results consistent to the standard star calibration at the 0.03~mag level.

Optical and near-infrared image reduction and photometry was performed using standard IRAF tasks \citep{tod93}. For each frame a model of the point spread function (PSF) was constructed using brighter field stars, and fitted to the afterglow. The relatively large seeing between 2\arcsec~and 3\arcsec~together with the pixel scale of 0\farc{16} for \griz and 0\farc{60} for \JHK resulted in an excellent spatial sampling of the PSF with statistical fit errors of order 0.2~\% for \griz and 0.5~\% for \JHK. For consistency, we also performed standard aperture photometry with compatible results with respect to the reported PSF photometry. All data were corrected for a Galactic foreground reddening of $E_{B-V}$ = 0.012~mag in the direction of the burst \citep{sch98}.

\begin{figure}
\epsscale{0.95}
\plotone{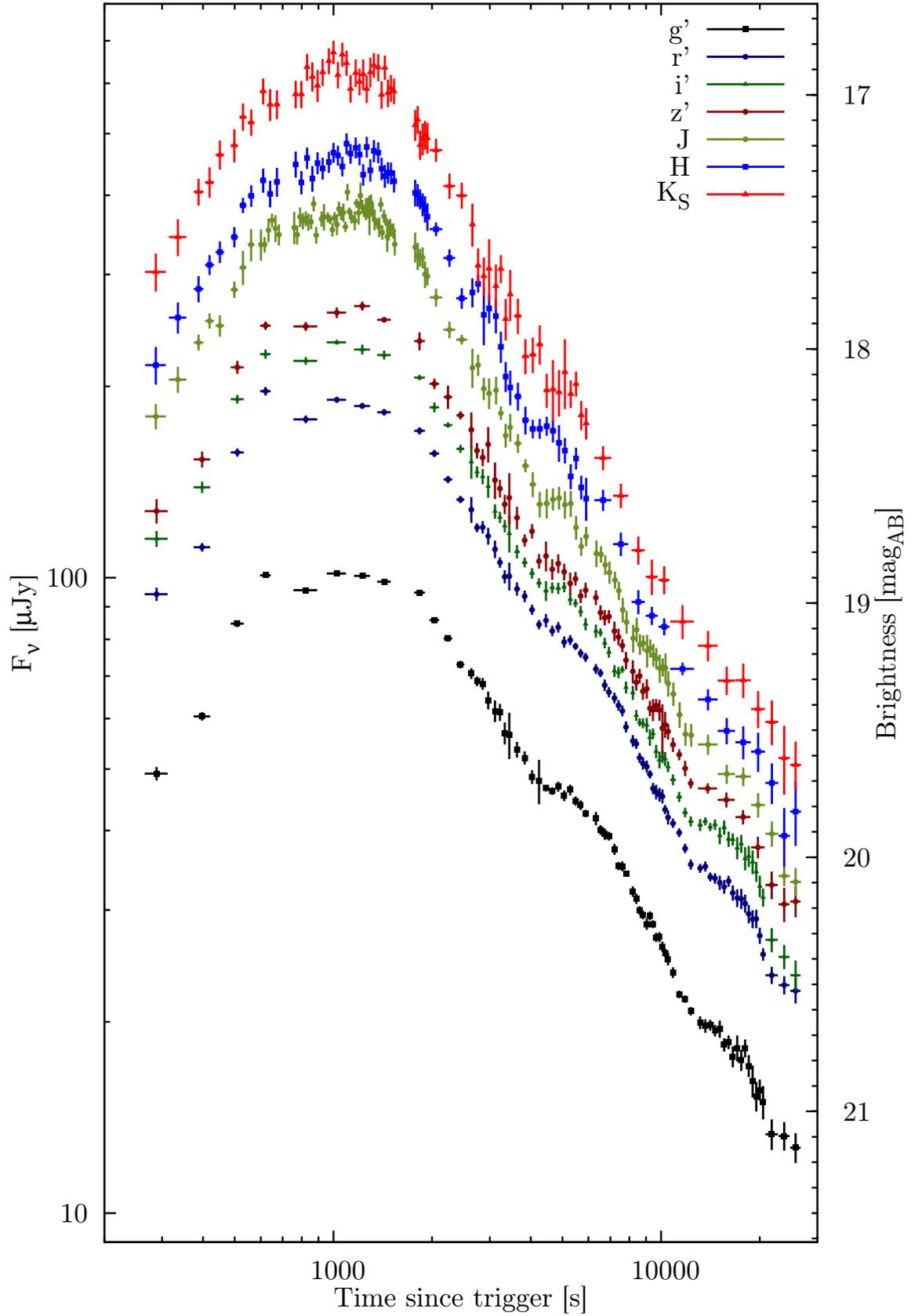}
\caption{GROND optical and NIR light curve of the afterglow of XRF~071031 taken between $\sim$4~min and 7~hours after the trigger.}
\label{picLight1}
\end{figure}

The stacking of individual images was done twice for different purposes. Firstly, all available data were used and individual frames were stacked until a statistical error in the PSF fit of around 0.1~mag was obtained. This resulted in 75 frames in each \gri, 51 in z$\arcmin$~, 118 in $J$ and 76 in each $H$ and $K_S$, yielding the multi-wavelength light curve shown in Fig.~\ref{picLight1}. Secondly, only NIR data simultaneous to the optical integrations were selected, excluding the frames which were taken during the \griz CCD read out. The resulting NIR image stacks were used to derive the optical to NIR spectral energy distribution (SED).

\section{Analysis}

\subsection{The optical/NIR light curve}

\begin{figure}
\epsscale{0.85}
\plotone{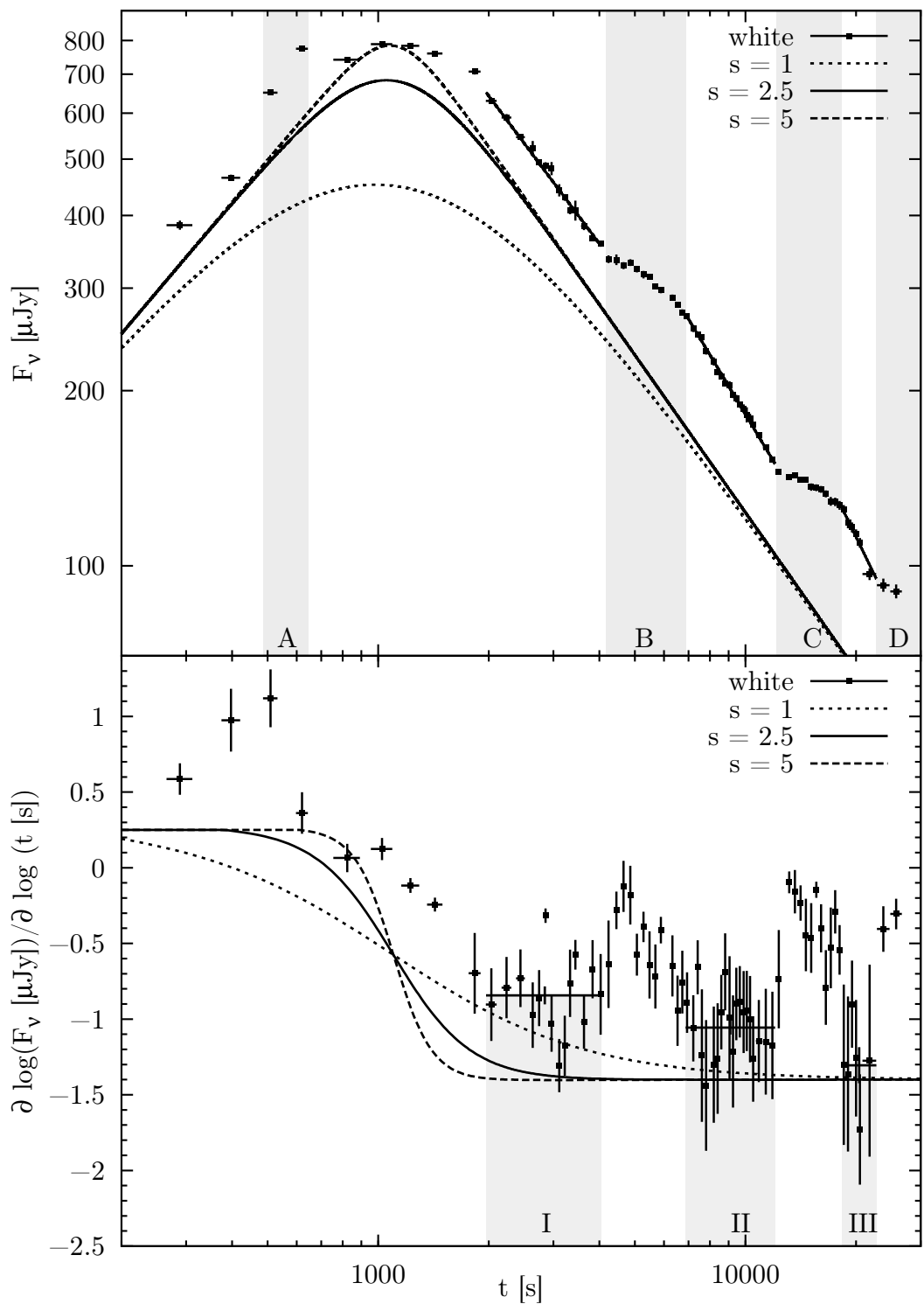}
\caption{White light curve of the afterglow of XRF~071031 in the upper panel. A numerical differentiation of the data in the native log-log scale is shown in the lower panel. Regions of bumpiness are denoted with capital letters (upper panel), and the steeping of the power law in Roman numbers (lower panel). The effect of a changing smoothness parameter s of the turnover is indicated by different lines in both panels.}
\label{picLight2}
\end{figure}

The multicolour light curve of XRF~071031 is complex and not described by smoothly connected power laws alone (Fig.~\ref{picLight1}). Evident in all colors is the initial increase in brightness up to $\sim$1~ks, which smoothly turns over to a generic power law decline with superimposed variations. In order to study small scale irregularities at the highest possible signal-to-noise, a white-light curve was derived by combining all \griz CCD data (Fig.~\ref{picLight2}, upper panel), which are exactly synchronous by hardware setup. The NIR bands are excluded from this process due to the intrinsically larger photometric error which would serve only to increase the uncertainties in the summed data. 

In order to better visualize the afterglow light curve, it is also presented differentiated in its native log-log scale, i.e. $\delta$(log(F$_{\nu}$))/$\delta$(log(t)), which directly represents the local power law decay index $\alpha$(t) (Fig.~\ref{picLight2}, lower panel). On a log-log scale, the first derivative of a power law is a constant, so for a smoothly connected power law rise and decay, one would expect a positive constant at early, turning into a negative at later times. Changes in the power law index and deviations from the decay are clearly visible in this representation. 

Without imposing an a priori model of the afterglow, Fig.~\ref{picLight2} convincingly demonstrates that the overall trend of the light curve is well described by two smoothly connected power laws as introduced in \citet{beu99}. In both panels of Fig.~\ref{picLight2}, however, the deviations from a Beuermann-like power law rise and decay are clearly apparent. There are two features, which require either a superimposed component, or a different parametrization of the intrinsic afterglow. The first and most obvious is the additional emission component in regions A,B,C,D shown in the upper panel of Fig.~\ref{picLight2}. This extra emission requires features intrinsic to the source or its environment to produce the observed flux excess with respect to the power law. The second is the steepening of the power law decline after a rebrightening from region I over II to III in the lower panel of Fig.~\ref{picLight2}. However, there is so much variability within the light curve that the underlying afterglow cannot be established with high certainty. 

The early rising component in the optical bands might be related to the deceleration of the FS by the circumburst medium, which happens when the swept up medium efficiently decelerate the ejecta. From the time of the light curve peak, the initial bulk Lorentz factor of the outflow $\Gamma_0$ can be constrained. Using the formalism outlined by \citet{sar99}, \citet{pan00} and \citet{mol07}, $\Gamma_0$ is estimated to $\approx 90 \left (\frac{E_{53}}{\eta_{0.2}A^{*}}\right)^{1/4}$ in a wind shaped circumburst medium and to $\approx 200 \left (\frac{E_{53}}{\eta_{0.2}n}\right)^{1/8}$ for an ISM type environment, with a weak dependence on the uncertain parameters $A^{*}$ being the normalized wind density, $E_{53}$ the isotropic-equivalent energy released in $\gamma$-rays in $10^{53}$~erg, $\eta_{0.2}$ the radiative transfer efficiency normalized to 0.2 and $n$ the ISM density in cm$^{-3}$. The slow rise with a power law index of $\sim$0.7 suggests a wind like environment, which would only be consistent with the closure relations for a very hard electron index \citep{dai01} of p$\sim$1.6 and $\nu_c<\nu$ using the spectral and temporal slopes in the late afterglow light curve $\alpha_{X,o}\sim$1 and $\beta_X$=0.8$\pm$0.1. The classical closure relations \citep[e.g.][]{zha04}, however, would favour an ISM environment in the slow cooling case with $\nu_m<\nu<\nu_c$ and a more canonical value of p$\sim$2.6.

Alternatively, the initial rise could be the result of a structured outflow seen off-axis \citep[e.g.][]{pan98}. In the case of previous fast and slowly rising afterglow light curves, \citet{pan08} find an anticorrelation of peak flux in the R band $F_{\nu,R}$ and peak time $t_p$. K-correcting the afterglow to $z$ = 2 to match the previous sample, we find that the optical/NIR light curve of XRF~071031 fits very well into this anticorrelation. In this interpretation the slow rise would hint on a smooth angular structure of the outflow \citep{pan08}.

Chromaticity around peak brightness was tested by comparing the optical/NIR SED before and after the total maximum. Apart from changes in the spectral index which can be attributed to the emergence of the bumps (see Sec. \ref{col} and Fig.~\ref{picSpec}), there is no evidence for a change in the spectrum before and after the light curve peak. The time of the light curve maximum is not correlated with energy and all bands peak at a similar time within the measurement uncertainties. Such evolution would be expected if the main peak was caused by cooling of the ejecta after the prompt emission resulting in the shift of the characteristic synchrotron frequency $\nu_m$ into the optical bands \citep[e.g.][]{zia08}. In addition, a moving $\nu_m$ through the optical bands is expected to cause a strong change from a positive to negative spectral index \citep[e.g.][]{sar98, gra02}. Neither effect is observed.

After correcting for Galactic foreground extinction, the SED from g$\arcmin$ to $K_S$ was fit with a power law and dust reddening templates in the host at $z$=2.692. Using extinction templates from the Milky Way (MW), Large- (LMC) and Small Magellanic Cloud (SMC) and supernovae induced dust \citep{pei92, maio04, stra07} no signatures of dust extinction in the GRB host are evident down to 1$\sigma$ confidence limits of $A_{\rm V}^{\rm host} <$ 0.06~mag (MW),  $A_{\rm V}^{\rm host} <$ 0.07~mag (LMC),  $A_{\rm V}^{\rm host} <$ 0.05~mag (SMC and SNe induced). Therefore the effect of dust reddening is considered as negligible in the following analysis. 
The deviation from a power law SED in the GROND g$\arcmin$ band is consistent with Lyman-$\alpha$ absorption in the GRB host at $z$=2.692.

\subsection{The X-ray afterglow light curve}

\begin{figure}
\epsscale{0.85}
\plotone{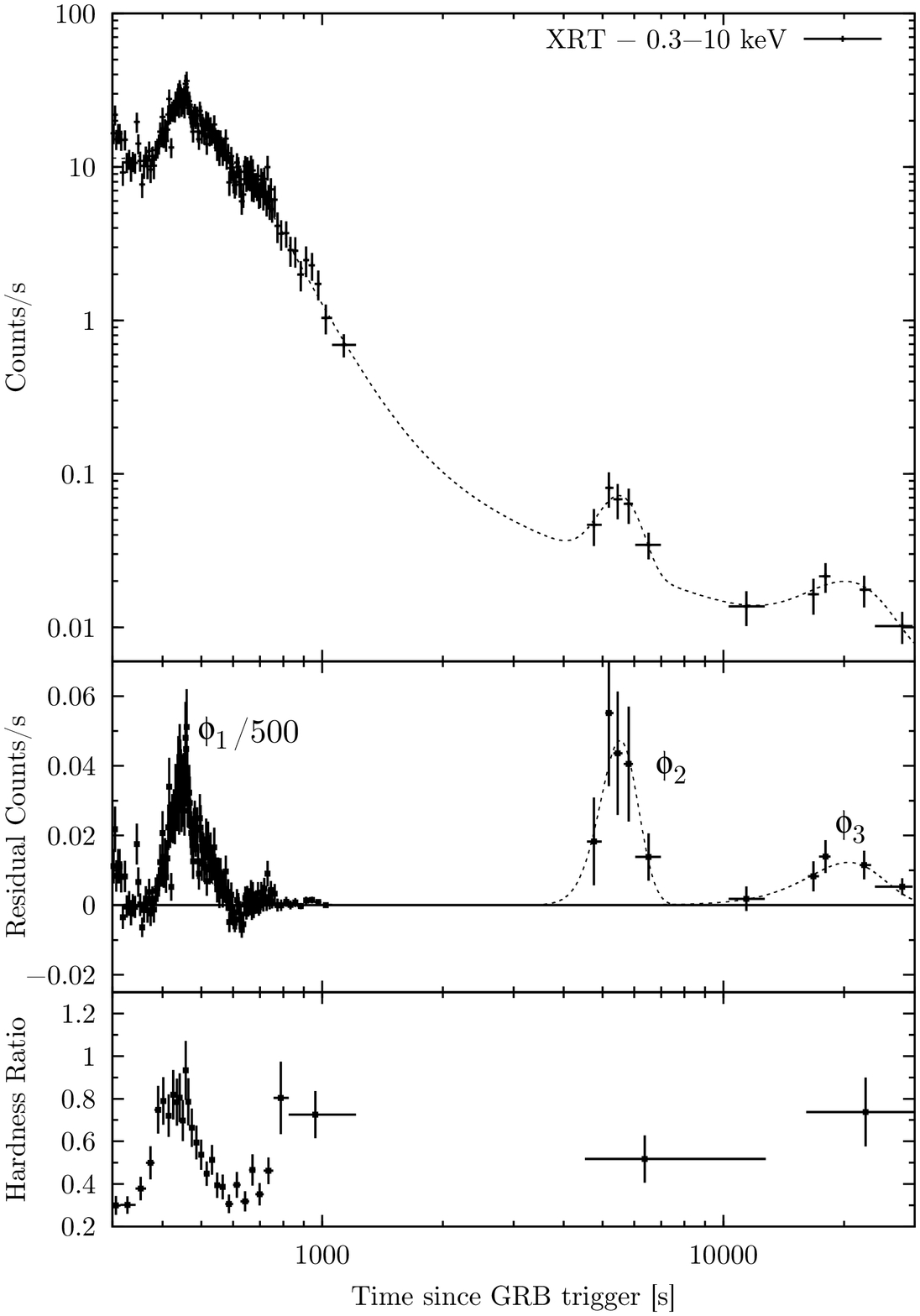}
\caption{XRT light curve of the afterglow of XRF~071031 in the time domain simultaneous to the GROND data (top panel). In the middle panel the residuals to the smoothly connected power law fit, as well as their modeling using Gaussians. The residuals of the first flare have been scaled by 1/500. The bottom panel shows the hardness ratio of the 1.5-10~keV versus 0.3-1.5~keV band.}
\label{picLight3}
\end{figure}

In addition to the variable and densely sampled light curve in the GROND filter bands, the X-ray afterglow is bright and well covered by XRT observations. Similar to what is seen in the optical bands, the X-ray data show strong variability, and the underlying afterglow is poorly constrained. After excluding the very early data, where there are no GROND observations (t-T$_0<$ 300~s), we fitted the remaining data using a similar procedure as used for the optical bands with a combination of a smoothly connected power laws. Remarkably, the obtained late power law index $\alpha_{\rm X}$=0.99$\pm$0.12 is compatible with the best fit from the GROND data $\alpha_{\rm o}$=0.97$\pm$0.06, providing additional evidence that the applied model fitting traces the underlying power law decay of the afterglow reasonably well. The XRT light curve and the residuals to the power law fits are shown in Fig.~\ref{picLight3}. After flaring episodes, the XRT light curve drops back to the power law, consistent with the rebrightenings observed in previous GRBs or XRFs \citep[e.g.][]{bur05b, rom06a}. All XRT light curve data have been obtained from the \textit{Swift} XRT light curve online repository \citep{eva07}. 

\subsection{The bumps}

Both X-ray and optical/NIR data show significant variations from the typical power law profiles at very early times. While this is observed in nearly 50\% of all X-ray afterglows \citep[e.g.][]{nou06}, this is rarely seen so early in an optical light curve starting $\sim$400~s after the trigger. 

\begin{figure}
\epsscale{0.85}
\plotone{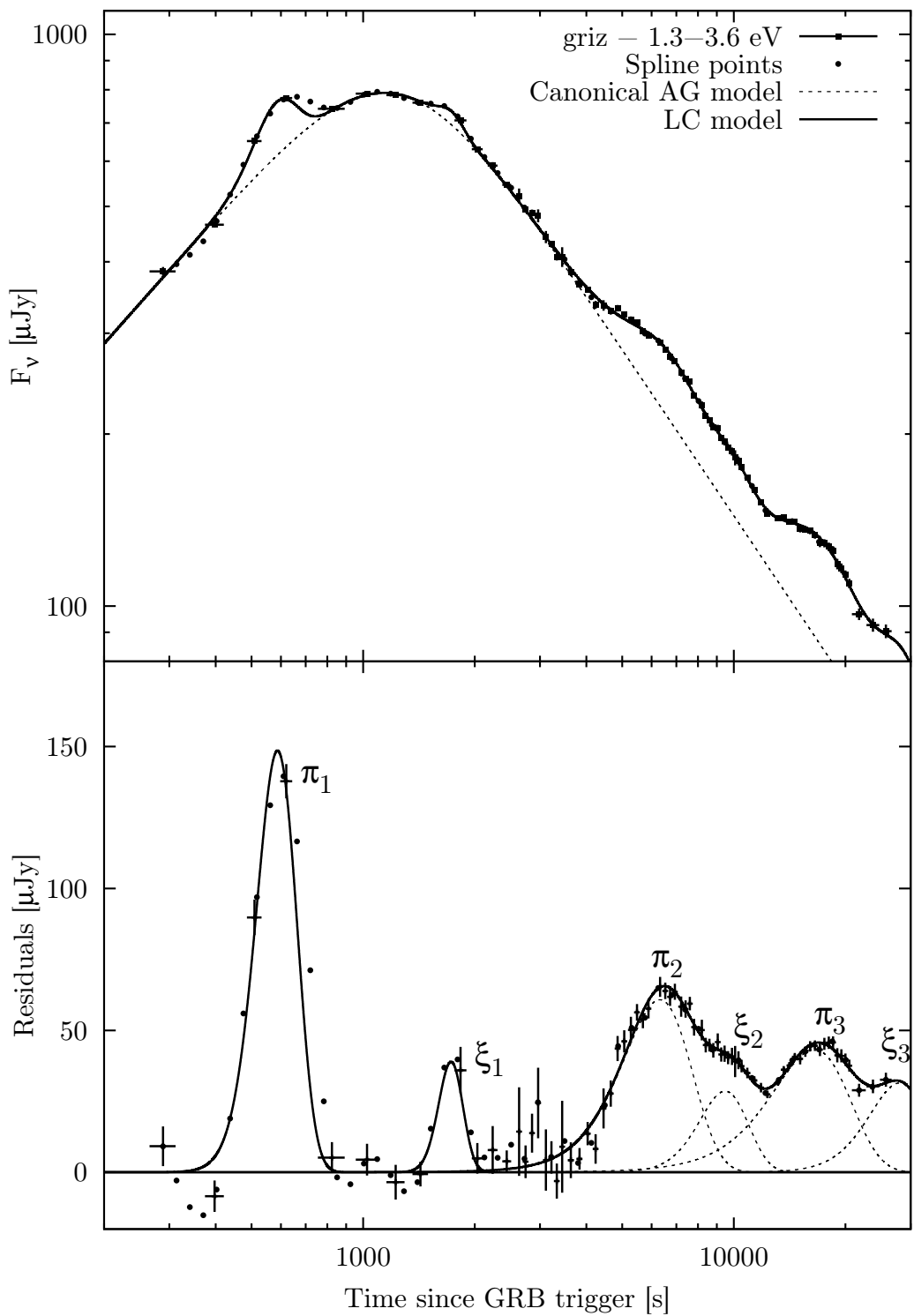}
\caption{White light curve of the afterglow of XRF~071031 (upper panel, shown are the data and the cubic spline interpolation with equally spaced nodes in log(t)). The data were fitted using the sum of a smoothly connected power law for the canonical afterglow (dashed line) and Gaussian profiles to account for the evident flux excess (solid line). In the lower panel the residuals to the smoothly connected power law, as well their modeling by six Gaussians.}
\label{picLight4}
\end{figure}

We fitted the combined white light data with a canonical afterglow rise and decay with superimposed Gaussian profiles to account for the variations in the light curve. While this may not be the true physical model, it provides a good fit to the data, and represents the morphology of the bumps with adequate accuracy. For the very early optical data, where the time sampling of the light curve is naturally sparse, we used a cubic spline interpolation with equally spaced nodes in the native log-log scale of the afterglow to constrain the fit. The data suggest the existence of three major and three small bumps: The three brighter ones, $\pi_1$ to $\pi_3$, which peak at 0.6~ks, 6~ks and 18~ks, and three fainter ones, $\xi_1$ to $\xi_3$, all shown in the lower panel of Fig.~\ref{picLight4}. The first faint bump is only indicated by one data point and the last peak is not sampled by the observations due to the break of dawn. The fit shown in Fig.~\ref{picLight4} has a $\chi^2_{red}$ = 1.05 with 52 degrees of freedom (dof). The best fit parameters are presented in Tab.~\ref{tab3}.

The typical timescale of variation for the optical bumps is $\langle\Delta$T/T$\rangle$ = 0.35 $\pm$ 0.13, where $\Delta$T is calculated as the full width half maximum of the Gaussian, and have a relative flux increase with respect to the underlying afterglow of $\langle\Delta$F/F$\rangle$ = 0.29 $\pm$ 0.18. All bumps, which are reasonably sampled by our observations, have a change in the slope $\delta \alpha$ between 0.5 and 0.9, which can be directly deduced from the lower panel of Fig.~\ref{picLight2}. 

Due to the faintness of the source, the late X-ray afterglow light curve is not equally well sampled as the optical and a detailed analysis is not possible in late epochs. There is, however, evidence for three X-ray flares $\phi_1$ to $\phi_3$ between 350 and 25000~s with synchronous optical coverage. To derive the morphology of the flares in the XRT light curve, we use a similar approach as for the optical bands for a direct comparison: An underlying continuum fitted with power laws and superimposed Gaussians to account for the evident flux excess which yields a reduced $\chi^2_{red}$ of 1.15 for 154 degrees of freedom. The best fit is shown in Fig.~\ref{picLight3} and the corresponding parameters are reported in Tab.~\ref{tab4}. 

The X-ray flares are much stronger $\langle\Delta$F/F$\rangle$ = 1.28 $\pm$ 0.28 with respect to the underlying afterglow than the optical bumps. Comparing against a statistical sample of previous X-ray flares \citep{chi06} shows that the flares observed in XRF~071031 populate a similar phasespace region of $\Delta$T/T versus $\Delta$F/F and thus resemble the morphology of previous flares.

\subsection{Spectral evolution}
\label{col}

\begin{figure}
\includegraphics[angle=270, width=0.98\columnwidth]{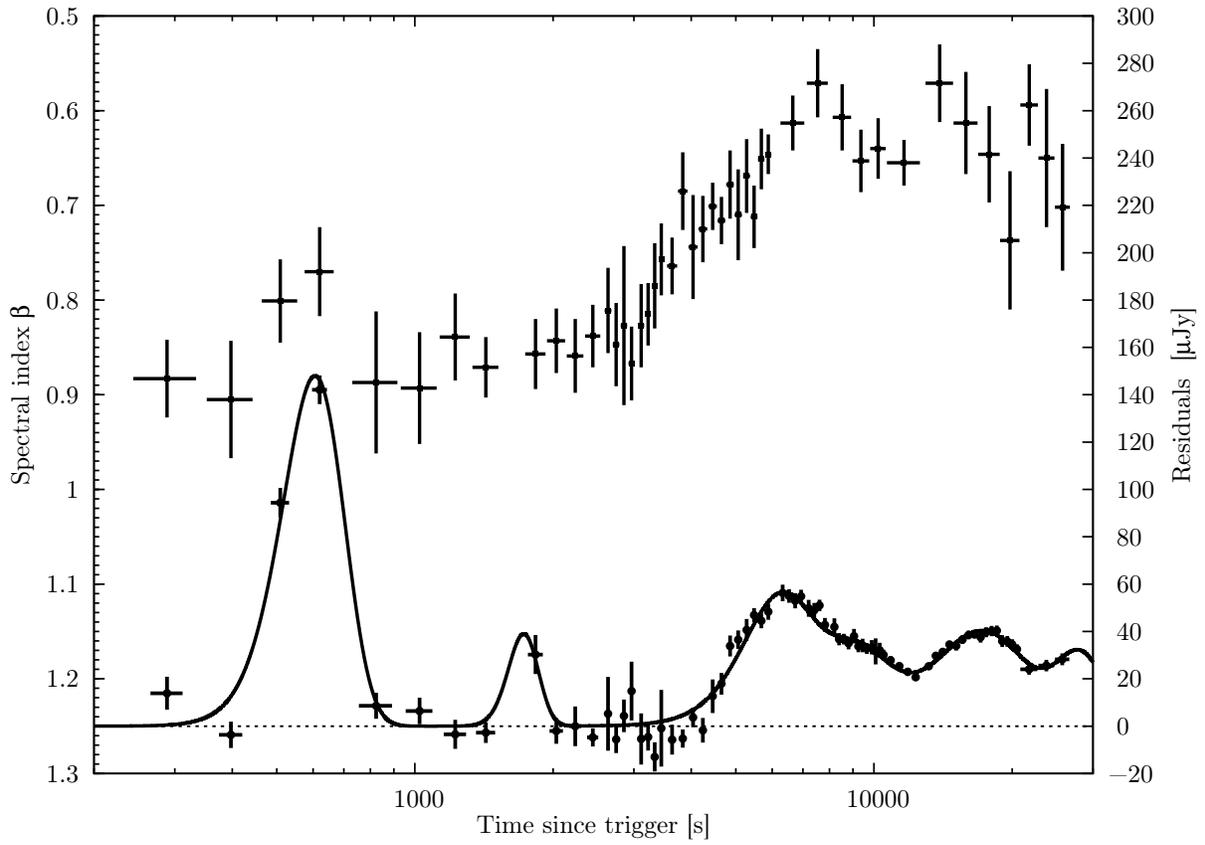}
\caption{Color evolution of the optical/NIR afterglow of XRF~071031. A direct comparison of the spectral index $\beta$ (upper points - left y axis) against the residuals to a fit of a canonical afterglow light curve model (lower points - right y axis). The plot clearly indicates a correlation of the spectral hardening and the rebrightening episodes in the light curve.}
\label{picSpec}
\end{figure}

The spectral index $\beta$, where $F_{\nu} \propto \nu^{-\beta}$ of the optical/NIR SED is observed to evolve with time (Fig.~\ref{picSpec}). Remarkably, the chromatic evolution is correlated with the residuals of the data against the light curve fits.
Applying a standard statistical correlation analysis yields a correlation coefficient of $\sim$-0.73, and thus a null hypothesis probability of $\sim 10^{-6}$.

The correlation of spectral hardening and bumpiness suggests, that this is an intrinsic feature of the emission component in the bumps, rather than the afterglow itself. In the ISM model for example, the cooling frequency $\nu_c$ moving through the optical bands, would identify itself by a spectral softening \citep{sar98}, in contrast to the observations. A hardening of the spectrum would be expected in a wind like environment \citep{gra02} with a change in the spectral index of 0.5. The observed change (see Fig.~\ref{picSpec}) is $\sim$0.2 and therefore not compatible with a cooling break passing the optical bands also in the wind model.


In addition to the observed chromatic evolution in the optical bands, also the X-ray data show strong spectral changes throughout the observations. This is already indicated by the evolving hardness ratio of the two XRT bands (1.5-10~keV over 0.3-1.5~keV), shown in the lowest panel of Fig.~\ref{picLight3}. 

This evolution becomes more evident, when including individual early flares from Fig.~\ref{picXRT} into the analysis. For each flare in Tab.~\ref{xrays}, a spectrum was extracted and fitted by single and broken power law models. The rest frame column density $N_{\rm H}$ at $z$=2.692 was obtained by fitting the late photon counting data and found to be consistent with zero within a 1$\sigma$ confidence level. Combined with the negligible $A_{\rm V}$ from the optical data and assuming a constant $N_{\rm H}$ in the burst environment, the intrinsic $N_{\rm H}$ is neglected in the spectral fits. There is strong evidence that the flares are better modeled by broken rather than single power laws as shown in Tab.~\ref{xrays}, which is similar to flares seen in e.g. GRB~051117A \citep{goa07}, GRB~050713A \citep{gue07} and 061121 \citep{pag07}. 

\begin{figure}
\includegraphics[angle=270, width=0.98\columnwidth]{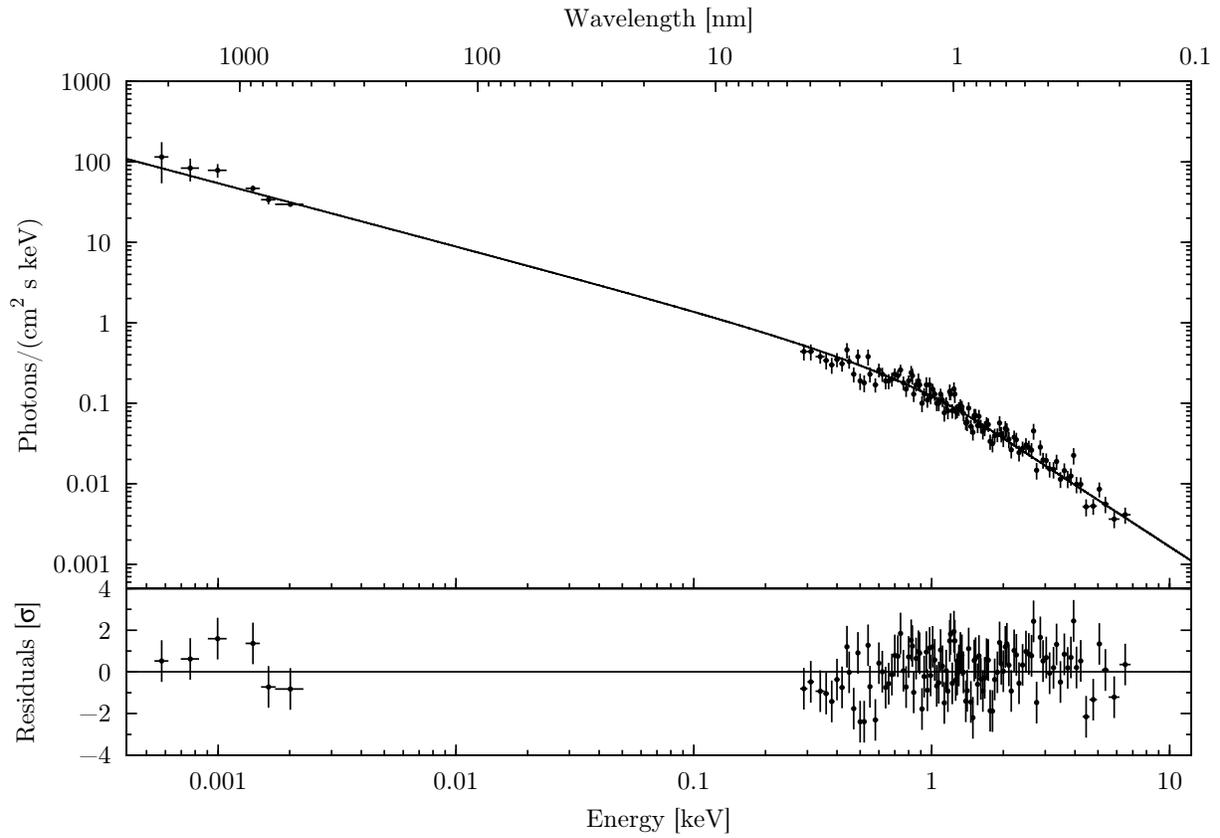}
\caption{Broadband spectrum of the first flare $\pi_1$ with simultaneous optical and X-ray coverage. The data are fitted by a Band function with E$_{\rm P}$=1.79$\pm$0.59 keV, $\alpha$=$-$0.78$\pm$0.03 and $\beta$=$-$1.92$^{+0.11}_{-0.17}$.}
\label{FigBBsed}
\end{figure}

From the X-ray data alone, there is evidence for a break in the spectrum in the 1~keV range for the early flares. This is consistent with the result of \citet{but07}, who find that the peak energy of the flare spectrum E$_{\rm P}$ crosses the X-ray bands on a typical timescale of 10$^2$ to 10$^4$ s. Combining the excess flux in the optical bands for $\pi_1$ and the X-ray spectrum in $\phi_1$, the spectrum can be constrained over a broad energy range. For the optical bands the dominant emission process is FS emission even at early times, and the afterglow model fitting was used to disentangle the different components. In this way estimates of the flux attributed to the flare component can be obtained. As shown in Fig.~\ref{FigBBsed}, the broadband spectrum of the first flare is reasonably well ($\chi^2$=129 with 114 d.o.f) described by a Band function \citep{ban93} with an E$_{\rm P}$ of 1.79 $\pm$ 0.59 keV and a very reasonable set of parameters $\alpha$=$-$0.78$\pm$0.03 and $\beta$=$-$1.92$^{+0.11}_{-0.17}$ as compared to the BATSE sample \citep{kan06}. 

We caution that this fit implicitly assumes that the excess emission seen in the optical bands is correlated with the X-ray flare and the applied model of the underlying FS emission traces the afterglow reasonably well. As the fundamental shape of the afterglow can be different then the empirical Beuermann profile, this might introduce significant systematic errors in the analysis. Additionally, an underestimated or even varying column density $N_{\rm H}$, would change the soft X-ray absorption and thus the overall broadband and X-ray fits.

For the later flares the different components attributed to FS and flare emission cannot be deconvolved with high certainty. The X-ray data are faint and affected by the underlying afterglow, and the excess flux in the optical bands is strongly dependent on the parameters of the light curve fitting. In particular, there is a strong ambiguity between the sharpness of the break and the light curve decay. Given the large uncertainties, the optical colour attributed to later bumps is compatible with the slope of the lower energy part of the Band function found for the first flare. E$_{\rm P}$, however, can no longer be constrained, but it is interesting to speculate that a similar Band function with an E$_{\rm P}$ between the optical and X-ray bands would account for the observed excess fluxes.

\section{Discussion}

\subsection{The likely cause of the flares}

Previously, bumps or flares in optical afterglow light curves have been reproduced using either a superimposed reverse shock component for early flares, inhomogeneities in the circumburst medium \citep[e.g.][]{wan00} or the angular distribution of the energy in the jet (patchy shell model, \citealp[e.g.][]{kum00}) or late energy injection by refreshed shocks \citep[e.g.][]{ree98} for later flares. However, a clear discrimination in the individual previous cases was not possible due to the lack of simultaneous coverage in broad wavelength ranges. 

Although we cannot completely rule out a reverse shock component for the first bump, it seems likely that it is produced by the same mechanism as the later ones. After subtracting the rising power law, the decline of the first bump can be fitted with a power law of index $\sim$-7. This would be surprisingly fast for a reverse shock, which is expected to decline with $\sim t^{-2}$ in the basic fireball model \citep[e.g.][]{nak04} and not faster than $\sim t^{-3}$ for more complicated models \citep{kob00}. The fact that the morphology is comparable to the later bumps additionally hints on a common origin. Therefore we try to account for all bumps with as few assumptions as possible, thus searching for a phenomenological explanation for all bumps observed.

Recently, \citet{nak07} found, that density jumps in the circumburst medium cannot account for the majority of fluctuations previously observed in GRB afterglow light curves. Only large contrasts in the circumburst medium density are able to produce bumps with a change in the temporal power law decay index $\delta\alpha\approx$1 in the light curve with a long transition time, which scales linearly with contrast. Thus, signatures of inhomogeneities in the circumburst medium in the optical light curve are expected to be smoother to what is observed. 

A jet with inhomogeneities in the angular energy distribution produces episodic bumps when the cone of a relativistically beamed patch enters the field of the view of the observer \citep{zha06}. Similar to refreshed shocks, these patches inject additional energy into the blastwave, the afterglow emission is boosted to a higher level and resumes the same power law index as before the bump \citep{zha02}. A characteristic for rebrightenings due to patchy shells or refreshed shocks consequentially is a step like afterglow light curve. Given the steepening of the power law post bump (Fig.~\ref{picLight2}) this scenario seems inconsistent with the bumps observed in XRF~071031. We note however, that the refreshed shock scenario can produce rebrightenings on relatively short timescales under certain conditions, where the light curve drops back to the initial decay \citep[e.g.][]{gui07}.

%
%
Although the morphology of the  light curve and optical bumps in the GROND data might be explained within the framework of variable external density or energy dissipation in the FS, both the spectral evolution as well as the correlation with the X-ray data argue for an independent origin and a second emission component. A hardening of the spectral index for flares in an optical light curve is not unprecedented \citep{gre08b}, and suggests a different emission than the generic afterglow FS. In addition, the first pronounced bump in the optical light curve is already observed during the rise of the afterglow at 600~s, when the apparent onset of the FS just started.

If all previously observed X-ray flares are due to the same physical process, it is very likely that they do not originate from external shocks that give rise to the afterglow emission, but from late time internal shocks \citep[e.g.][]{bur05b, zha06, chi06, but07}. Thus, X-ray flares seem to be produced by a similar mechanism as the prompt $\gamma$-rays, which are also caused by internal shocks in the standard model \citep{ree92}. Detailed analysis for GRB~050820A \citep{ves06} showed that the optical emission contemporaneous with the prompt phase can be explained as the superposition of forward shock and emission correlated with the $\gamma$-rays. Similarly for XRF~071031, after subtracting the dominating FS in the optical bands, the flare spectrum from NIR to X-rays is well described with a Band function.


Remarkably, the optical bumps show features which have been previously observed in X-ray flares: a hardening of their spectra \citep[e.g.][]{bur05b, but07, fal07, goa07} and a correlation of the duration with the time where the bump occurs, i.e. a roughly constant $\Delta$T/T \citep{chi06, koc07}. They are however, less pronounced than typical X-ray flares. Temporal analysis of the early data shows, that the optical bump peaks significantly later than the X-ray flare. A hard to soft evolution of E$_{\rm P}$ which is found in the majority of all bright flares where a detailed spectral analysis is possible \citep[e.g.][]{bur05b, rom06b, perri07, fal07} provides a natural explanation for the time difference between the flare in the X-ray and optical wavelength range. Spectral lags and a broadening towards lower energies have been observed in a number of previous X-ray flares and the prompt emission \citep[e.g.][]{nor96, rom06a, perri07}.

Based on the light curve fitting, the observed peak of the early optical flare is delayed by $\tau\sim$ 130~s compared to the X-rays, which corresponds to 35~s in the bursts rest frame. As the temporal coverage of the optical light curve is sparse in the early time frame, this delay is strongly dependent on the assumed functional form of the afterglow and flare morphology, but in any case it is significantly longer than what is typically observed as spectral lags in the prompt phase. Typical values for prompt lags range from slightly negative (i.e. soft preceding hard bands) to several seconds for long lag GRBs \citep[e.g.][]{nor00, geh06, smb08, fol08}. The observed time difference in the case of XRF~071031, however, is based on entirely different energy ranges. In particular, the difference between hard and soft energy bands is around a factor of 10$^3$ for X-rays versus optical, while it is $\sim$10 for BAT channels.

\begin{figure}
\includegraphics[angle=270, width=0.98\columnwidth]{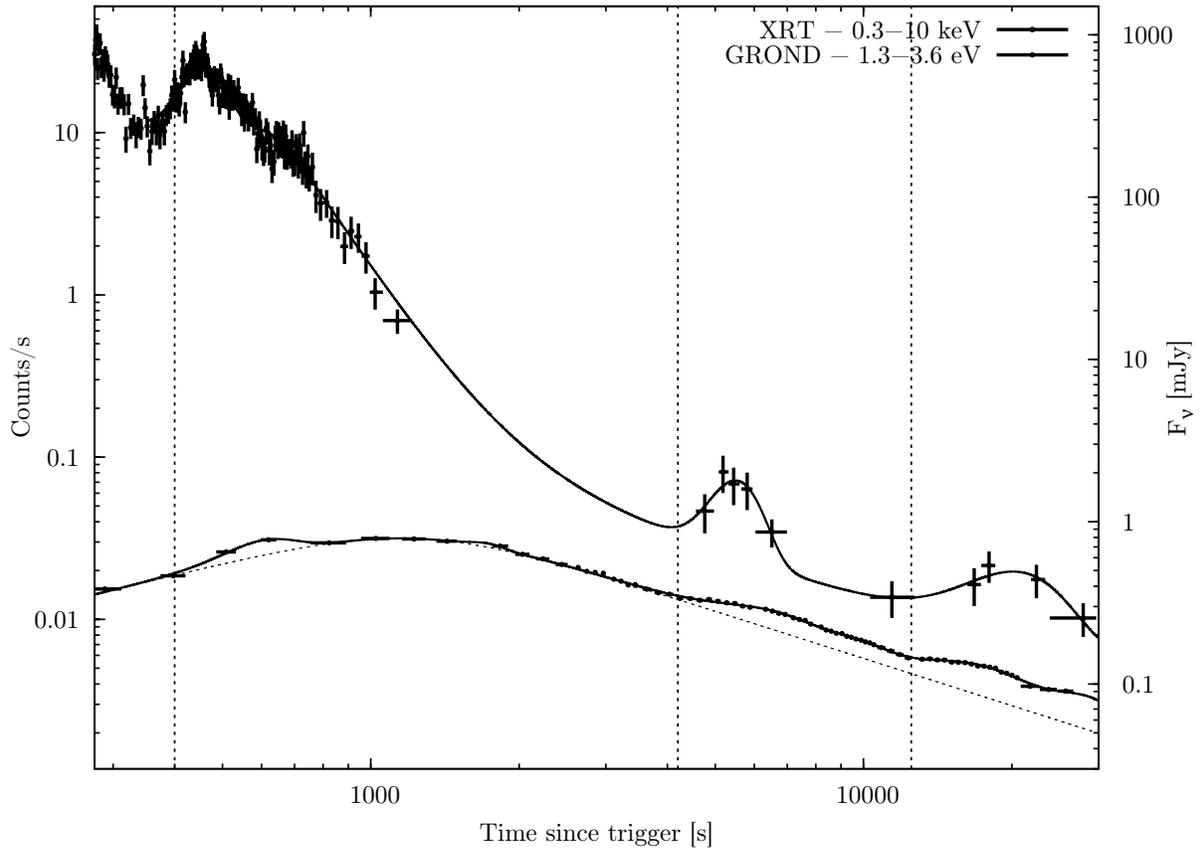}
\caption{Combined plot of XRT light curve (crosses, left y axis) and GROND white light (dots, right y axis). The vertical dashed lines indicate the emergence of the bumps from the underlying optical afterglow emission.}
\label{picLight5}
\end{figure}

Combining bump morphology, colour evolution, broadband spectrum and the temporal connection to the X-ray data as shown in Fig.~\ref{picLight5}, the most likely origin of the bumps in the optical/NIR light curve of XRF~071031 is the same as in the X-rays, namely the soft tail of emission correlated with late internal shocks. 

\subsection{Paucity of detection of correlated early optical bumps and X-ray flares}

While a number of bursts have shown considerable variability in the optical bands during X-ray flares \citep[e.g.][]{pan06,boe06,pag07,mal07}, the early optical light curve is generally not as strongly affected as the X-rays by flaring episodes. The UVOT and ground based robotic telescopes with their fast response provide a good sample of optical observations simultaneous with the XRT light curve. If the first optical bump and the first X-ray flare in XRF~071031 are correlated as suggested by the data, the signature of a typical X-ray flare in the optical light curve can be approximated with the ratio of absolute optical to X-ray peak flux obtained for XRF~071031.


GRB~060607A for example, showed significant X-ray flaring very early in the X-ray light curve with only marginally correlated optical emission \citep{zia08}. However, the afterglow was very bright, reaching $\sim$14.3~mag in the r band \citep{nys07}. If the flare was described with a similar spectrum as in XRF~071031, the expected optical counterpart of the brightest X-ray flare is estimated to have a peak amplitude of $\sim$200 $\mu$Jy and could only be resolved with continuous photometric monitoring with a precision of at least 0.03~mag. Thus, if the emission in the flares is not strong enough with respect to the FS, a bright afterglow can easily outshine flare signatures in the optical bands even for very bright X-ray flares. In fact, GRB~061121 \citep{pag07}, had simultaneous coverage with BAT, XRT and UVOT of the prompt emission with a peak flux density in the X-rays of $\sim$15 mJy. \citet{pag07} find, that most of the flux is emitted in the $\gamma$-ray and X-ray bands, while the UVOT data only showed a relatively small increase in brightness. 

\section{Conclusions}

The detailed light curve and spectral evolution of the afterglow of XRF~071031 showed several remarkable features. A general trend of an achromatically rising and falling optical/NIR afterglow, consistent with a number of previous bursts where an early increase in brightness was reported. The achromatic turnover from rise to decay suggests the apparent onset of the FS as its origin, either due to the increase in radiating electrons in the pre-deceleration phase or a structured outflow seen off axis. In the latter case the increase in brightness is caused by the deceleration of the shock, so that the relativistically beamed cone widens and gradually enters the sight line of the observer.

Superimposed onto the afterglow continuum emission are bumps which have a harder SED and appear in similar epochs in the optical/NIR as well as in the X-ray bands. Based on the GROND data collected for XRF~071031 we conclude that the variation in the light curve are most probably the optical counterparts of X-ray flares, and therefore directly connect variability in an optical light curve with the emergence of X-ray flares. Although emission from external shocks or a combination of different effects cannot be completely ruled out, an internal origin seems to nicely account for the majority of observations: The light curve shape and in particular the morphology of the bumps, the spectral hardening in the optical SED, the observed decrease in $E_{\rm P}$ from the prompt emission to the flares and the broadband flare spectrum from NIR to X-rays. The spectral similarities to the prompt phase strengthen the picture of X-rays flares as later and softer examples of the prompt emission due to internal shocks. This connection provides additional evidence that inner engine activity may last or be revived on a timescale of hours or days at least for some bursts.

\acknowledgments
We thank the referee for a very helpful report and comments and D. H. Hartmann for discussion. This research was supported by the DFG cluster of excellence 'Origin and Structure of the Universe'. A.R. and S.K. acknowledge support by DFG grant Kl 766/11-3. Part of the funding for GROND (both hardware as well as personnel) was generously granted from the Leibniz-Prize to Prof. G. Hasinger (DFG grant HA 1850/28-1). This work made use of data supplied by the UK Swift Science Data Centre at the University of Leicester.


\begin{deluxetable}{cccccccc}
\tabletypesize{\scriptsize}
\tablecaption{Secondary standards in the GRB field in the GROND filter bands from 1 to 10
\label{tabSecStan}}
\tablewidth{0pt}
\tablehead{
\colhead {Ra/Dec} & \colhead {g$\arcmin$} & \colhead {r$\arcmin$} & \colhead {i$\arcmin$} & \colhead {z$\arcmin$} & \colhead{J} & \colhead{H} & \colhead{K$_S$} \\
Deg [J2000] &  mag & mag & mag & mag & mag & mag & mag}
\startdata
6.4098/-58.0831 & 15.44 $\pm$ 0.02 & 14.55 $\pm$ 0.02 & 14.28 $\pm$ 0.02 & 14.13 $\pm$ 0.02 & 13.09 $\pm$ 0.04 & 12.55 $\pm$ 0.04 & 12.50 $\pm$ 0.06 \\
6.4186/-58.0723 & 16.32 $\pm$ 0.02 & 15.73 $\pm$ 0.02 & 15.54 $\pm$ 0.02 & 15.43 $\pm$ 0.02 & 14.46 $\pm$ 0.04 & 14.07 $\pm$ 0.04 & 14.00 $\pm$ 0.06 \\
6.3933/-58.0662 & 17.84 $\pm$ 0.03 & 17.32 $\pm$ 0.03 & 17.18 $\pm$ 0.03 & 17.10 $\pm$ 0.03 & 16.18 $\pm$ 0.04 & 15.94 $\pm$ 0.06 & 15.65 $\pm$ 0.07 \\
6.3672/-58.0536 & 20.71 $\pm$ 0.05 & 19.33 $\pm$ 0.03 & 18.53 $\pm$ 0.03 & 18.13 $\pm$ 0.03 & 16.86 $\pm$ 0.05 & 16.41 $\pm$ 0.07 & 16.28 $\pm$ 0.08 \\
6.4579/-58.0492 & 20.03 $\pm$ 0.04 & 18.90 $\pm$ 0.03 & 18.49 $\pm$ 0.03 & 18.31 $\pm$ 0.03 & 17.07 $\pm$ 0.05 & 16.68 $\pm$ 0.07 & 16.45 $\pm$ 0.08 \\
6.3663/-58.0782 & 20.09 $\pm$ 0.04 & 18.62 $\pm$ 0.03 & 17.60 $\pm$ 0.03 & 17.13 $\pm$ 0.03 & 15.80 $\pm$ 0.04 & 15.30 $\pm$ 0.05 & 14.92 $\pm$ 0.06 \\
6.3751/-58.0631 & 21.26 $\pm$ 0.05 & 20.13 $\pm$ 0.04 & 19.72 $\pm$ 0.04 & 19.52 $\pm$ 0.04 & 18.29 $\pm$ 0.07 & 17.87 $\pm$ 0.08 & 17.79 $\pm$ 0.10 \\
6.4021/-58.0496 & 22.72 $\pm$ 0.10 & 20.85 $\pm$ 0.05 & 20.06 $\pm$ 0.05 & 19.65 $\pm$ 0.05 & 17.94 $\pm$ 0.06 & 17.35 $\pm$ 0.07 & 16.11 $\pm$ 0.08 \\
6.4117/-58.0575 & 22.49 $\pm$ 0.08 & 21.21 $\pm$ 0.06 & 20.76 $\pm$ 0.06 & 20.51 $\pm$ 0.07 & 18.56 $\pm$ 0.07 & 17.78 $\pm$ 0.08 & 16.86 $\pm$ 0.10 \\
6.4375/-58.0485 & 21.28 $\pm$ 0.05 & 20.77 $\pm$ 0.05 & 20.61 $\pm$ 0.05 & 20.59 $\pm$ 0.07 & 19.29 $\pm$ 0.09 & 18.91 $\pm$ 0.10 & 17.98 $\pm$ 0.15 \\
\hline
\enddata
\end{deluxetable}
\begin{deluxetable}{cccccc}
\tabletypesize{\scriptsize}
\tablecaption{Parametrization of the excess flux in the GROND bands using Gaussians
\label{tab3}}
\tablewidth{0pt}
\tablehead{\colhead{Bump} & \colhead{T$_{mid}$ [s]} & \colhead{$\Delta$T/2 [s]} & \colhead{Amplitude [$\mu$Jy]} & \colhead{$\Delta$T/T$_{mid}$} & \colhead{$\Delta$F/F}}
\startdata
$\pi_1$ & 587 $\pm$ 12 & 84 $\pm$ 18 & 149 $\pm$ 5 & 0.29 $\pm$ 0.06 & 0.24 $\pm$ 0.01 \\
$\xi_1$ & 1726 $\pm$ 6 & 130 $\pm$ 4 & 39 $\pm$ 5 & 0.15 $\pm$ 0.01 & 0.06 $\pm$ 0.01 \\
$\pi_2$ & 6302 $\pm$ 156 & 1648 $\pm$ 136 &  60 $\pm$ 5 & 0.52 $\pm$ 0.05 &  0.26 $\pm$ 0.02 \\
$\xi_2$ & 9439 $\pm$ 377 &  1461 $\pm$ 254  &  29 $\pm$ 4 & 0.31 $\pm$ 0.06 & 0.18 $\pm$ 0.02\\
$\pi_3$ & 16505 $\pm$ 135 & 4013 $\pm$ 263 & 43 $\pm$ 2 & 0.49 $\pm$ 0.03 & 0.46 $\pm$ 0.02\\
$\xi_3$ & 28000 &  5000 & 32 $\pm$ 4 & 0.36 & 0.55 $\pm$ 0.03\\
\hline
\enddata
\end{deluxetable}
\begin{deluxetable}{cccccc}
\tabletypesize{\scriptsize}
\tablecaption{Parametrization of the excess flux in the X-ray bands using Gaussians
\label{tab4}}
\tablewidth{0pt}
\tablehead{\colhead{Bump} & \colhead{T$_{mid}$ [s]} & \colhead{$\Delta$T/2 [s]} & \colhead{Amplitude [Counts/s]} & \colhead{$\Delta$T/T$_{mid}$} & \colhead{$\Delta$F/F}}
\startdata
$\phi_1$ & 459 $\pm$ 2 & 42 $\pm$ 2 & 14.5 $\pm$ 0.8 & 0.18 $\pm$ 0.01 & 1.36 $\pm$ 0.07 \\
$\phi_2$ & 5528$\pm$ 85 & 604 $\pm$ 114 & 0.047 $\pm$ 0.01 & 0.22 $\pm$ 0.04 & 1.20 $\pm$ 0.20  \\
$\phi_3$ & 20507 $\pm$ 952 & 4557 $\pm$ 1088 & 0.012 $\pm$ 0.002 & 0.44 $\pm$ 0.10 & 1.28 $\pm$ 0.13 \\
\hline
\enddata
\end{deluxetable}
\begin{deluxetable}{ccccccc}
\tabletypesize{\scriptsize}
\tablecaption{Spectral fits to the XRT flares using XSPEC. 
\label{xrays}}
\tablewidth{0pt}
\tablehead{
\colhead {Flare} & \colhead {Times [s]} &\colhead {Model} & \colhead {Photon Index 1} &  \colhead {Break energy [keV]} & \colhead {Photon Index 2} & \colhead {$\chi^2$/dof}}
\startdata
1  & 137 - 150 & Single & 1.23$^{+0.05}_{-0.05}$ & - & - & 323/131 \\
1  & 137 - 150 & Double & 0.72$^{+0.09}_{-0.09}$ & 1.84$^{+0.17}_{-0.36}$  & 1.85$^{+0.12}_{-0.19}$ & 157/129 \\
\hline
2  & 189 - 217 & Single & 1.72$^{+0.09}_{-0.05}$ & - & - & 180/85 \\
2  & 189 - 217 & Double & 1.14$^{+0.14}_{-0.15}$ & 1.26$^{+0.16}_{-0.13}$ & 2.20$^{+0.13}_{-0.11}$ & 89/83 \\
\hline
3  & 236 - 284 & Single & 1.87$^{+0.05}_{-0.05}$ & - & - & 146/74 \\
3  & 236 - 284 & Double & 1.23$^{+0.19}_{-0.18}$ & 1.07$^{+0.16}_{-0.12}$ & 2.39$^{+0.17}_{-0.14}$ & 67/72 \\
\hline
4 ($\Phi_1$) & 396 - 547 & Single & 1.64$^{+0.04}_{-0.04}$ & - & - & 172/110 \\
4 ($\Phi_1$) & 396 - 547 & Double & 1.26$^{+0.19}_{-0.32}$ & 0.98$^{+0.22}_{-0.20}$ & 1.88$^{+0.08}_{-0.08}$ & 120/108 \\
\hline
\enddata
\end{deluxetable}
\end{document}